\documentclass[12pt]{article}
\usepackage{latexsym,amsmath,amssymb,amsbsy,graphicx}
\usepackage[space]{cite} 

\begin{document}

\noindent\begin{minipage}{\textwidth}
\begin{center}

{\Large{On the possibility of the implicit  renormalization procedure for the Casimir energy }}\\[9pt]

\medskip\small

\textbf{A.\,I. Dubikovsky$^{1a}$, P.\,K. Silaev$^{2b}$}

\smallskip

\textit
{$^1$Department of Quantum Statistics and Field Theory,
Faculty of Physics, M.\,V.\,Lomonosov Moscow State University, Moscow 119991, Russia.}\\
\textit {$^2$Department of Quantum Theory and High Energy Physics,
Faculty of Physics, M.\,V.\,Lomonosov Moscow State University, Moscow 119991, Russia.}\\
\textit {E-mail: $^a$dubikovs@physics.msu.ru,
$^b$silaev@bog.msu.ru}

\medskip

\end{center}

{\parindent5mm We propose a procedure for the renormalization of Casimir energy, that is
based on the implicit versions of standard steps of  renormalization procedure --- regularization, subtraction and removing the regularization.
The proposed procedure is based on the calculation of
a set of convergent sums, which are related to the original divergent sum
for the non-renormalized Casimir energy. Then one constructs a  linear equations system,
that connects this set of convergent sums with the renormalized Casimir energy, which  is  a  solution 
to this equations system. This procedure slightly reduces the 
indeterminancy that arises in standard procedure when we choose  the particular regularization and the explicit form of counterterm. 
The proposed procedure is more efficient from the computational point of view than the standard one. It can be 
applied not only for systems with the explicit transcendental equation for the spectrum, but also for systems with the spectrum that can be obtained only numerically.\vspace{2pt}\par}

 \vspace {1pt} \par

\vspace{1pt}\par
\end{minipage}

\section*{Introduction}
\mbox{}\vspace{-\baselineskip}

Since the pioneering paper of H.\,B.\,G.\,Casimir \cite {casim}  a   large
number of various methods for regularizing and renormalizing of the Casimir energy was proposed. 
Let us recall some of them --- starting from the simplest one, that  was used in the first works including \cite {casim}. It applies an exponential cutoff function  introduced into expression for the energy. Then the minimal subtraction is performed, i.e.  all singular terms (terms that became singular when the cutoff parameter tends to zero) are dropped out. One can also use the Abel-Plan formula \cite {abel1, abel2} (this approach requires an explicit expression for the spectrum).  We can also mention more sophisticated methods: regularization with the zeta function \cite{dzeta1, dzeta2, milton, bordag}; procedure, based on the Green function 
\cite{lopez1, lopez2, lopez3, lopez4}; 
procedure, based on the coefficients  of the heat kernel  expansion \cite {heat1, heat2} (this method is in some sense  equivalent to the method with  zeta function), and many others. For trivial cases --- flat boundaries, separated bodies, etc., all the methods mentioned above  yield identical answers. And this result  can be easily explained, since in all these cases the renormalization   can be reduced to minimal subtraction, whereas the properties of  singularities for different regularizations appears to be more or less similar. In more complex cases the situation changes. For instance for the calculation of Casimir energy  in the ball \cite {heat2}, it turns out that in different regularizations the singularities have different properties, i.e.  terms that are regular in the zeta function regularization appears to be  singular in other regularizations.  Moreover, one can find at least three different answers for  the Casimir energy  in the bag model (obtained by different methods).  It should be noted that one of the answers \cite {bag3m, bag4m}, have an opposite sign in comparison with the other two, obtained in \cite {bag1p, bag2p} and in \cite {bag5pp}. We should also mention the problem of the dielectric ball. One can extend  the well-known solution for the infinitely thin conducting sphere \cite {shar1, shar2} to the case of a dielectric ball only by imposing certain conditions on the dielectric and magnetic permeability of the sphere and  of the  external medium \cite {uslov1, uslov2}. An alternative solution to this problem is possible in the limit of a small perturbation, when the relative permittivity is close to unity \cite {vozm1, vozm2, vozm3, vozm4}. Another method of choosing the normalization point for the Casimir energy in a ball allowing  to solve the problem of a dielectric ball (this method does not rely on the problem of an infinitely thin conducting sphere) was proposed in \cite {shar3, shar4}.

Recently, interest to the Casimir effect has increased significantly,  due to the development of new precision measurements methods ---
in particular, it appears to be possible to measure  Casimir force of the opposite sign, i.e. repulsive \cite {ottalke} (it arises in the case of a certain conditions imposed on the  dielectric permittivity of bodies and the external medium \cite {ottalkt1, ottalkt2}.
It has been  studied  the dependence of the Casimir force  between two bodies on their shape \cite {forma1, forma2}, and also the effects of interaction between a macroscopic and microscopic object (atom).
Therefore   problems mentioned above, that are not fully solved yet, become especially relevant. Strictly speaking the solution to these problems  reduces to the choice of the normalization point for  the Casimir energy or the Casimir pressure. Moreover, then one should give a proof  that the obtained answer does not depend on the choice of regularization (typically we should   confine ourselves by a certain class of regularizations).

So we can ask a question: is it possible to avoid this problem by an implicit renormalization procedure, i.e. procedure constructed in such a way, that all three standard steps  --- regularization, subtraction and removing of regularization appears to be implicit? One should recall, that the indeterminancy arises when we choose  the particular  regularization and when we choose the explicit form of counterterm. 
Definitely, in such implicit procedure both regularization and subtraction are  inevitably present, but they are implicit. So it is not necessary to  define a specific regularizing function (that belongs to a certain class of functions) and  specify an explicit counterterm form (that becomes singular when the regularization is removed), which then will be subtracted during the renormalization procedure.

In this paper we propose exactly such an implicit procedure.
It is in some sense a combination of well-known renormalization methods, adapted to the calculation  of the Casimir energy.
In the dispersion relation method, linear combinations constructed from the integrand function \cite {bmp, oa} are used to achieve the convergence of integrals, in procedure of  renormalizations in quantum field theory  differentiation on the  parameters of the theory (both usual differential operator and the operator of finite difference) are frequently used.  
But we apply these methods to a problem where  (in some sense)  there is no ``well-defined'' normalization point (for example, the standard normalization on the observed  values of the constants of the theory).

We shall illustrate the proposed procedure   by a simple example, that can  be easily solved by standard methods. This will allow us to verify that the proposed procedure actually leads to  a correct and well-known answer.
The choice of boundary conditions in this example is motivated by the following reasons: the imposed boundary conditions yield us a spectrum that is  similar to the spectrum in multidimensional problems with non-zero curvature of boundaries (for instance, in a ball). For such  spectra the logarithmic divergences in the Casimir energy arise not only due to the mass of the field (as it would be for zero boundary conditions), but also due to the intrinsic properties of the spectrum.

\section * {Simple one-dimensional example }
\mbox {} \vspace {- \baselineskip}

Let us consider the simplest example:  the scalar field   in  one-dimensional (1 + 1) space on the interval $ [0, a] $ with mixed boundary conditions with the standard lagrangian:
$$ L = {1 \over 2} (\partial_0 \varphi (t, x)) ^ 2- {1 \over 2} (\partial_1 \varphi (t, x)) ^ 2- {m ^ 2 \over 2} (\varphi (t, x)) ^ 2 $$
We choose the following boundary conditions:
$$ \varphi (t, 0) = 0, \qquad \partial_1 \varphi (t, a) + \lambda \varphi (t, a) = 0.
$$
For such a system there is no explicit analytic expression for the spectrum, so  the expression for the Casimir energy
\begin {equation}
E = {1 \over 2} \sum_ {n = 1} ^ \infty \sqrt {\mathstrut k_n ^ 2 + m ^ 2} \label {energy}
\end {equation}
contains the values $ k_n $, that are the roots of the transcendental equation
$$ k_n \cos (k_n a) + \lambda \sin (k_n a) = 0. $$

One of the standard renormalization procedures for such cases is the following \cite {bordag}:
we should perform  integration along the contour $C$ on the complex plane $ k $
$$ {1 \over 2 \pi i} \int \limits_C \; dk \; \sqrt {k ^ 2 + m ^ 2} \; {f '(k) \over f (k)} \; \Phi (k), $$
where $ f (k) $ is the equation for the roots  $ k_n $:
$$ f (k) \equiv \cos (k a) + \lambda \sin (k a) / k. $$
Here the function $ \Phi (k) $ is a regularizing function that makes integrals and sums finite,
and one can choose between  two possible contours $ C $  described below (the  choice of the contour will only affect the requirements imposed on the regularization function $ \Phi (k) $,  the final answers appears to be identical in both cases). The first contour lays along the imaginary axis shifted by an infinitesimal distance to the right (due to  square root  $ \sqrt {\mathstrut k ^ 2 + m ^ 2} $  two branching points are present at the imaginary axis, and  besides this  for the case $ \lambda <0 $ the function $f(k)$ possibly have a root on the imaginary axis --- the  ``discrete'' level). This contour is closed to the right by an arc of a semicircle with infinite radius.
On the other hand, the second  contour runs along the real axis, shifted upward to an infinitesimal distance (the roots $k_n$ of the function $ f $  provide the poles on the real axis). The contour is closed to the upward by an arc of a semicircle with infinite radius, bypassing  the cut along the imaginary axis from $ k = im $ to $ k \to i \infty $. For the first  contour the function $ \Phi $ must decrease rapidly when $ k $ tends to $ + \infty $ (as $ 1 / k ^ \gamma $, $ \gamma> 2 $) and should not have poles to the right of the imaginary axis.
So, even the simplest polynomial regularization $ \Phi (k) = 1 / (1+ \epsilon k) ^ \gamma $ (where $ \epsilon> 0 $, $ \gamma> 2 $) is quite acceptable, and the regularization can be removed if we set $ \epsilon \to 0 $). Similarly, for the second contour the  function $ \Phi $ should  have no poles above the real axis,  so  they can be placed on the negative part of the imaginary axis: $ \Phi (k) = 1 / (1 + i \epsilon k) ^ \gamma $.
In both cases  the integral over an arc of a semicircle of infinite radius is zero.
Further we  consider only the case of the first contour, for the second contour the consideration appears to be quite similar. 
If we take into account that all poles that correspond to the sum
(\ref {energy}) are located inside the contour  to the right of
the imaginary axis,  we obtain
$$ - {1 \over 2 \pi} \int \limits _ {- \infty} ^ {\infty} \; d \kappa \; \sqrt {(i \kappa + 0) ^ 2 + m ^ 2} \; {f'(i \kappa) \over f (i \kappa + 0)} \; \Phi (i \kappa) = \sum_ {n = 1} ^ \infty \sqrt {\mathstrut k_n ^ 2 + m ^ 2} \; \Phi (k_n).
$$

Now let us  turn to the renormalization. If we choose  the  normalization point
$ a \to \infty $, we obtain
$$
{F '(i \kappa) \over f (i \kappa)} \; {\vrule height 4 ex depth 2.5 ex} _ {\; \; a \to \infty} = {- ia \sinh (\kappa A) -ia \lambda \cosh (\kappa a) / \kappa + i \lambda \sinh (\kappa a) / \kappa ^ 2 \over \cosh (\kappa a) + \lambda \sinh (\kappa a ) / \kappa} \; {\vrule height 4 ex depth 2.5 ex} _ {\; \; a \to \infty} = $$ $$ = - i \, \hbox {sign} (\kappa) {a (1+ \lambda / | \kappa |) - \lambda / \kappa ^ 2 \over 1+ \lambda / | \kappa |}
$$
Consequently (for this choice of the normalization point) we subtract the terms that are proportional to the first and zero powers of $ a $. The remaining expression decreases exponentially with the increasing  $ | \kappa | $, so the regularization in renormalized expression can be removed:
$$
E^{(ren)}=-{1\over 2\pi }\int\limits_{-\infty}^{\infty} \; d\kappa \;\sqrt{(i\kappa+0)^2+m^2} \; \left[{f'(i\kappa+0)\over f(i\kappa+0)} - {f'(i \kappa+0)\over f(i \kappa+0)}\;{\;\vrule height 4 ex depth 2.5 ex}_{\;\;a\to\infty}  \right]
$$
If we take into account that the function $ f '/ f $ is odd, and  the square root has an opposite signs  on two intervals $ [im, i \infty] $ and $ [-im, -i \infty] $,   we can conclude that the 
contribution to the  integral from the interval $ [- im, im] $
disappears, whereas the contributions from the intervals $ [im, i \infty] $ and $ [-im, -i \infty] $ appears to be equal and so yield duplicated result.   
 Finally we obtain
\begin{equation}
E^{(ren)}(a)=-{1\over \pi }\int\limits_{m}^{\infty} \; d\kappa \;\sqrt{\kappa^2-m^2} \; { a-\lambda/(\kappa^2-\lambda^2)\over e^{2\kappa a}(\kappa+\lambda)/(\kappa-\lambda)+1
} \label{pereno}
\end{equation}
In this expression for negative $ \lambda $ can appear a non-integrable singularity caused by the complex root of function $f$ (it corresponds to the solution with  imaginary wave number --- ``discrete level'').
However, this situation implies the realization of nonphysical case of an exponentially ``exploding'' field, because not only the wave number $ \kappa_0 $ appears to be imaginary,
but also  the corresponding  frequency $ \omega_0 = \sqrt {\mathstrut - \kappa_0 ^ 2 + m ^ 2} $ is imaginary (provided that $m<\kappa_0$). 

It should be emphasized that due to our specific choice of renormalization point we have excluded the ``surface energy'' from the obtained result for energy. 
The ``surface energy'' are the contributions to energy that are associated with the boundaries of the interval $ x = 0 $ and $ x = a $.  In the expression for $ T_ {00} $, one can find the  contribution that depends on the coordinate $ x $ and isn't equal to zero only at the vicinity  of both boundaries  $ x = 0 $ and $ x = a $.  This surface energy obviously does not vanish for infinite interval length $ a \to \infty $. However, all  values that can be experimentally observed for our system (i.e.  the energy difference for different values of $ a $, or the force (``pressure'')  on the boundaries) do not depend on the surface energy.
Nevertheless, if necessary, this surface energy can be calculated by a computation, that is completely analogous to the computation outlined above. This computation yields an answer that varies in the range from $ 0 $ (for $ \lambda = 0 $) to $ -m / 4 $  (for $ \lambda \to \infty $) and does not depend on the length of the interval  $ a $.

The expression for the renormalized energy  decreases exponentially with increasing mass of the field (or with increasing  length of the interval  $ a $)  and is in complete agreement with the answers for forces (``pressures'') on both boundaries  $ x = 0 $ and $ x = a $ of the interval (we can find the answer for forces if we calculate the average value of $ T_ {11} $ in the vacuum state  instead of
 the average value of $ T_ {00} $).

\section * {Implicit renormalization procedure }
\mbox {} \vspace {- \baselineskip}

Let us ask a question whether it  is  possible  to avoid (in some sense) 
the standard renormalization procedure with regularization and a choice of the explicit form of counterterm.  In other words, is it possible to construct a procedure in which both standard steps in the renormalization turn out to be implicit? This procedure will deal only with convergent sums related to the Casimir energy, and do not use the regularization function $ \Phi (k) $ explicitly. It should be emphasized, however, that it is necessary to impose a regularization when we   establish the relation between  the constructed convergent sum  and the initial divergent sum for the Casimir energy. 

But during the process of calculation of the Casimir energy the regularization function $ \Phi (k) $ does not appear at all. 

At first glance, the construction of such a procedure should be equivalent to excluding terms, that are proportional to the first and zero power of the interval length  $a$.
But this is true only for a regularized expression, the procedure of subtraction and   the procedure of removal of regularization do not commute.
  
In divergent sum
\begin {equation}
E (a) = \sum_ {n = 1} ^ \infty \sqrt {\mathstrut (k_n (a)) ^ 2 + m ^ 2} \label {nereg}
\end {equation}
the linearly divergent terms are present and they are proportional to $ \pi n / a $.
Also we find the logarithmically divergent terms  (related to the mass of the field), they are
proportional to $am^2/(2\pi n)$. And, finally, there are the logarithmically divergent terms,
arising due to the boundary condition of the third kind, that are proportional to $ \lambda / (\pi n) $.

If one performs  a ``minimum subtraction'', i.e. if one subtracts from the sum (\ref {nereg}) all the divergent terms listed above, then the resulting finite answer will contain terms proportional to $ \log a $ and $ a \log a $. These terms 
increase with increasing $ a $ and should be excluded in the process of renormalization.

Provided the sum is regularized, the contribution of the linearly divergent terms will be proportional to $ 1 / \varepsilon ^ 2 $, and the contribution of  logarithmically divergent terms will be proportional to $ \log (\varepsilon) $  (here $ \varepsilon $ is the regularization parameter). Both divergent terms, as mentioned above, are proportional to the zero and first powers of $ a $.  But  in addition, as a result of regularization,   the additional finite contribution appears, which  completely compensates  the terms proportional to $ a \log a $.
Therefore, in a regularized expression, it is sufficient to exclude terms of the form $ a ^ 0 $ and $ a ^ 1 $, while in the unregularized one we should exclude terms proportional to
 $ a ^ {- 1} $, $ a ^ 0 $, $ a ^ 1 $, $ \log a $ and $ a \log a $.
 
Now let us consider the set of sums (\ref {nereg}) for 6 different $ a_i $. Without loss of generality
we can set $ a_i = a_0 + i $, where $ i = 0 \ldots 5 $. It is possible to consider another set $ a_i $, but this simplest set turns out to be convenient for subsequent analysis. Let us construct  a linear combination
\begin {equation}
E_6 (a_0) = \sum_ {i = 0} ^ 5 c_i (a_0) E (a_i). \label {konsum}
\end {equation}
Obviously, by means of  appropriate choice of  constants $ c_i $, we can exclude
from the expression (\ref {konsum}) not only divergent terms (proportional to $ a ^ {- 1} $, $ a ^ {0} $ and $ a ^ {1} $), but also terms that should be eliminated in the course of renormalization (i.e. terms, proportional to $ \log a $ and $ a \log a $).
It is sufficient to impose  the following conditions on constants $ c_i $:
$$
\sum_ {i = 0} ^ 5 c_i (a_0) / a_i = \sum_ {i = 0} ^ 5 c_i (a_0) = \sum_ {i = 0} ^ 5 c_i (a_0) a_i = \sum_ {i = 0} ^ 5 c_i (a_0) \log a_i =
$$
\begin {equation}
  = \sum_ {i = 0} ^ 5 c_i (a_0) a_i \log a_i = 0. \label {urav}
\end {equation}
Taking into account  the linear independence of all 5 functions included in this system of equations, we can find a solution to equation (\ref {urav}) if we express values  $ c_1 \ldots c_5 $  as a function of arbitrary $ c_0 $.
Up to this point we dealt with divergent expressions (\ref {nereg}) and (\ref {konsum}). From now on the  precise meaning of these expressions should be specified by the regularization. 
But if values  $ c_i $  satisfy equations (\ref {urav}), then the sum 
\begin {equation}
 E_6 (a_0) = \sum_ {n = 1} ^ \infty \sum_ {i = 0} ^ 5 c_i (a_0) \sqrt {\mathstrut (k_n (a_i)) ^ 2 + m ^ 2}, \label {shod}
\end {equation}
appears to be convergent and it is not necessary to introduce an explicit  regularization function $ \Phi $ into it.

The practical significance of the introduced value  $ E_6 (a_0) $ is that it decreases rapidly (exponentially) with increasing $ a_0 $.  In order to prove this statement it is sufficient to  recall the 
consideration that permits us to obtain the expression for renormalized energy (\ref {pereno}). We can regularize the sum (\ref {shod}), and after the regularization it is possible to change the order of summation by indices  $ n $ and $ i $.  The 
``renormalization''  in the double sum will be performed automatically, due  to the conditions (\ref {urav}) imposed on the constants $ c_i $. After this implicit subtraction the regularization can be removed, and we obtain
$$ E_6 (a_0) = \sum_ {i = 0} ^ 5 c_i (a_0) E ^ {(ren)} (a_i). $$
For each $a_i$  the term $ E ^ {(ren)} (a_i) $ decreases exponentially with increasing $ a_0 $, so the entire sum will also decrease exponentially with increasing $ a_0 $.

 Now let us consider the set $ a_i = a_0 + i $, where $ i = 0 \ldots \infty $.
Suppose that for each of these $ a_i $ we find $ E_6 (a_ {i}) $ in exactly the same way as for $ a_0 $. (We should remind once more that the uniform step $ h = 1 $ in the set $ a_i $ is chosen only for the sake of simplicity).
Then we obtain the following infinite system of equations
\begin {equation} \sum_ {i = 0} ^ 5 c_i (a_j) E ^ {(ren)} (a_ {j + i}) = E_6 (a_j), \label {slu2} \end {equation}
where in  the right-hand side of equations we find values $ E_6 (a_ {j}) $  ($ j = 0, \ldots \infty $), that  we can  calculate  in accordance with (\ref{shod}). Therefore  the unknown quantities
$ E ^ {(ren)} (a_ {i}) $ are the exact solution to this infinite system of equations.

Now let us turn to the problem of solving an infinite system of equations. We should  emphasize once more that the system  (\ref {slu})  have a practical significance  only if the
 right-hand side (and, consequently, $E ^ {(ren)} (a_j)$) rapidly decreases  with  increasing 
  index $j$. In our simple example this decrease appears to be exponential, so for each given accuracy $ \delta $ one can specify an index  $ i_0 $ such that $ E ^ {(ren)} (a_i) <\delta $ for each $ i> i_0 $.
So we can write the finite system of linear equations, that coincides with (\ref {slu2}), but index $j$ is limited by the cutoff index $i_0$:  
\begin {equation} \sum_ {i = 0} ^ 5 c_i (a_j) E ^ {(ren)} (a_ {j + i}) = E_6 (a_j), \label {slu} \end {equation}
 $ j = 0 \ldots i_0 $. The right-hand side of this system $ E_6 (a_j) $ can be calculated in a straightforward way (see (\ref{shod})), so approximate values of 
 $ E ^ {(ren)} (a_ {j}) $ for $ j = 0 \ldots i_0 $ are the solutions to finite $i_0\times i_0$ system (\ref{slu}). Values  $ E ^ {(ren)} (a_ {j}) $ for $ j = i_0 + 1, \ldots i_0 + 5 $ are less than $ \delta $, and we assume them equal to zero.
But if we want to increase the efficiency of computations,  it is convenient to calculate the characteristic decrease rate $ \xi \equiv E_6 (a_ {j + 1}) / E_6 (a_ {j}) $
and set $ E ^ {(ren)} (a_ {i_0 + j}) = E ^ {(ren)} (a_ {i_0}) \xi ^ j $ for $ j = 1, \ldots 5 $ instead of precise zero. This  allows one  to significantly decrease the value of cutoff index $ i_0 $ (size of finite matrix), that provides the convergence. We suppose that convergence is achieved  for a given $ i_0 $ if the answer for $ E ^ {(ren)} (a_ {0}) $ does not change (within the specified accuracy) with the further increase of the index  $ i_0 $.

One can suppose that we should give a proof that  for a fixed $ a_0 $ and $ i_0 $ the system (\ref {slu})  is not singular. However,  it is not necessary.  Indeed, the determinant of the linear system (\ref {slu}) is a function of the step $ h $ between successive values $ a_i $.
(For the  set  $a_i$  that we are considering  the step $ h $ is  equal to 1). If the determinant turns out to be zero for this step value, one can change it  and the determinant will differ from zero.  From the computational  point of view, the matrix singularity  appears when the eigenvalues of  the matrix (\ref {slu})  differ substantially in magnitude, so we can make
system (\ref {slu}) regular (from the computational point of view), simply by changing the step $ h $.

All the consideration  given above  was verified by a direct computation for various $ m $, $ a $, and $ \lambda $.
Firstly, we have verified that the relations (\ref {slu}) are satisfied for all $ j $, if the right-hand side of $ E_6 (a_ {j}) $ are calculated in accordance with
(\ref {shod}), and  the values $ E ^ {(ren)} (a_ {j}) $ are calculated according to (\ref {pereno}).
Secondly, it was verified that the straightforward solution of the system (\ref {slu}) permits us to find the value $ E ^ {(ren)} (a_ {0}) $, that coincides (within the specified accuracy) with the value calculated according to (\ref {pereno}). The higher is the value of the cutoff index $ i_0 $, the higher is the resulting accuracy.

Now, just for an illustration, we write out values $ E_6 (a_ {i}) $ (calculated in accordance with (\ref{shod})) and the renormalized energy $ E ^ {(ren)} (a_ {i}) $ (obtained as solution to (\ref {slu})) for parameter values $ \lambda = 3 $ and $ m = 1 $. The difference between the solution to (\ref {slu}) and $ E ^ {(ren)} (a_ {i}) $ obtained by straightforward calculation  (\ref {pereno})  for the cutoff index $ i_0 = 14 $  appears to be less then  $ 10 ^ {-12} $:
 
\vskip 4 mm 
 {
 
\scriptsize

\begin{tabular}{|l|l|l|   l|l|l| l|l|l|l|l|} \hline 
$a_i$ &
1& 2& 3& 4 &
 5& 6&  7\\\hline
$E_6(a_{i})$ &
 $ -8.79\cdot 10^{-5}$&
 $-4.45\cdot 10^{-5}$&
 $-7.97\cdot 10^{-6}$&
$-1.20\cdot 10^{-6}$&
$-1.70\cdot 10^{-7}$&
$-2.34\cdot 10^{-8}$&
$-3.18\cdot 10^{-9}$
\\\hline
$E^{(ren)}(a_{i})$&
$-1.83\cdot 10^{-2}$&
$-1.79\cdot 10^{-3}$&
$-2.00\cdot 10^{-4}$&
$-2.35\cdot 10^{-5}$&
$-2.85\cdot 10^{-6}$&
$-3.53\cdot 10^{-7}$&
$-4.42\cdot 10^{-8}$
\\\hline
\end{tabular}

}

\vskip 4 mm

We also present the analogous results,
obtained for the values $ \lambda = 1 $ and $ m = 1 $. The  discrepancy in  $ E ^ {(ren)} (a_ {i}) $
 does not exceed $ 10 ^ {-15} $ for the same value of the cutoff index $ i_0 = 14 $:

\vskip 4 mm 
{
 
\scriptsize 

\begin{tabular}{|l|l|l|   l|l|l| l|l|l|l|l|} \hline 
$a_i$ &
1& 2& 3& 4 & 5& 6&  7 \\\hline
$E_6(a_{i})$ &
$1.25\cdot 10^{-4}$&
$1.03\cdot 10^{-5}$&
$9.71\cdot 10^{-7}$&
$9.89\cdot 10^{-8}$&
$1.06\cdot 10^{-8}$&
$1.17\cdot 10^{-9}$&
$1.32\cdot 10^{-10}$\\\hline
$E^{(ren)}(a_{i})$&
$5.07\cdot 10^{-3}$&
$2.40\cdot 10^{-4}$&
$1.75\cdot 10^{-5}$&
$1.53\cdot 10^{-6}$&
$1.47\cdot 10^{-7}$&
$1.51\cdot 10^{-8}$&
$1.62\cdot 10^{-9}$\\ \hline
\end{tabular}

}

\vskip 4 mm

Finally, we give the results,
obtained for the values $ \lambda = 1/3 $ and $ m = 1 $. The difference between the solution to (\ref{slu}) and value of  $ E ^ {(ren)} (a_ {i}) $
 obtained according to (\ref {pereno}) is  less then $ 10 ^ {-13} $ for the same value of the cutoff  index $ i_0 = 14 $:

\vskip 4 mm 
{
 
\scriptsize 

\begin{tabular}{|l|l|l|   l|l|l| l|l|l|l|l|} \hline 
$a_i$ &
1& 2& 3& 4 &
 5& 6&  7\\\hline
$E_6(a_{i})$ &
$2.40\cdot 10^{-4}$&
$5.92\cdot 10^{-5}$&
$9.42\cdot 10^{-6}$&
$1.35\cdot 10^{-6}$&
$1.86\cdot 10^{-7}$&
$2.52\cdot 10^{-8}$&
$3.38\cdot 10^{-9}$\\\hline
$E^{(ren)}(a_{i})$&
$2.51\cdot 10^{-2}$&
$2.15\cdot 10^{-3}$&
$2.26\cdot 10^{-4}$&
$2.58\cdot 10^{-5}$&
$3.08\cdot 10^{-6}$&
$3.76\cdot 10^{-7}$&
$4.67\cdot 10^{-8}$
\\ \hline
\end{tabular}

}

\vskip 4 mm 
  
It should be noted, that  for the fixed values of $ m $ and $ \lambda $, one can find all values  $ E ^ {(ren)} (a_ {i}) $ for the whole set of $ a_i $  by one calculation --- as a solution to the system (\ref {slu}).

\section*{Conclusions}
\mbox{}\vspace{-\baselineskip}

Therefore we can conclude that the proposed  procedure for calculating the Casimir energy (at least for the simplest case considered) yields the  well-known correct answer. Both standard steps of renormalization procedure --- regularization and subtraction ---  are implicit, and it is not necessary to specify explicit form of regularization and explicit form of subtraction term. 
It allows us (in some sense) to decrease the ambiguity  associated with the choice of a  regularization and the choice of the  counterterm expression. 
Definitely, the regularization of divergent sums and the subtraction of divergent and nonphysical terms still occur, but both actions appear to be implicit.  In the calculation process  the regularizing function does not appear at all  and there is no explicit subtraction.    The proposed procedure  deals  with convergent sums only, and these finite sums allow us to find the ``renormalized''  Casimir energy.
But it should be stressed that we need standard regularization procedure to specify the relation between our convergent sums and the initial divergent sum for the Casimir energy.
 
From the computational point of view, the proposed procedure appears to be sufficiently effective (more efficient than the standard one), provided the convergent sums decrease rapidly with increasing interval length (in our example this decrease turns out to be exponential). The only disadvantage of our procedure is the necessity to perform calculations with an increased number of precision digits  to provide the desired accuracy for the solution of  the resulting  linear equations system.
 
It should be noted that the proposed procedure can be applied to any
divergent sum. Particularly, it can be applied in those cases when not only there is no explicit expression for the spectrum, but also there is no a transcendental equation for the system spectrum
 (this is a quite typical situation for multidimensional problems that are not quasi-one-dimensional).
But in order  to apply the proposed procedure in this case, we need to know not only  the expressions for divergent terms (this problem can be  solved easily, because the asymptotic expression for the spectrum usually can be calculated for most systems), but also expressions for nonphysical terms (in our case these expressions are proportional to $a\log a$ and $\log a$).
One can suspect that the classification of the terms into ``physical terms'' and   ``unphysical terms'' is an arbitrary action corresponding to the choice of the counterterm in the standard procedure. But this conclusion is not quite correct. Firstly, the proposed procedure completely eliminates the dependence on the regularization, since regularization is implicit. Secondly, the subtraction is also implicit, so our procedure eliminates the  	uncertainty,   that appears when we specify the explicit form of the counterterm  subtracted in the standard renormalization procedure.  (Generally speaking,   indeterminancy  arise due to the following circumstance:  one can add any ``physical'' term  to the counterterm and thus obtain alternative answer --- finite, physical, but different.) 
So we may hope that  the proposed method 
 brings us closer to the solution for the ``dielectric ball problem'' and other problems mentioned in the introduction.
However, it should be noted, that the finding of expressions for nonphysical terms in the general case is a separate problem requiring additional consideration.


\end {document}